\pdfoutput=1	%
\documentclass[9pt,twocolumn,twoside]{pnas-new}
\setboolean{displaywatermark}{false}

\templatetype{pnasresearcharticle} %

\title{A Hardware Platform for Efficient Multi-Modal Sensing with Adaptive Approximation}

\author[1]{Phillip Stanley-Marbell}
\author[2]{Martin Rinard} 

\affil[1]{Department of Engineering, University of Cambridge, Cambridge CB3 0FA, UK}
\affil[2]{Computer Science and Artificial Intelligence Laboratory (CSAIL), MIT, Cambridge Massachusetts, MA 02139, USA}

\leadauthor{Stanley-Marbell} 

\authorcontributions{PSM formulated the problem, designed the hardware, and performed the measurements. Both authors contributed to the writing.}
\correspondingauthor{\textsuperscript{1}To whom correspondence should be addressed. E-mail: {phillip.stanley-marbell@eng.cam.ac.uk}}

\vspace{-0.1in}

\usepackage{etoolbox}	%
\newtoggle{warpTightFormatting}
\newtoggle{warpAcmTecs}
\togglefalse{warpTightFormatting}
\usepackage{subfig}
\usepackage{xspace}
\usepackage{bm}
\usepackage{graphicx}
\usepackage{grffile}	%
\usepackage{textcomp}
\usepackage{color}
\usepackage{colortbl}
\usepackage{booktabs}

\usepackage{pifont}				%
\usepackage[scaled]{beramono}
\usepackage[T1]{fontenc}

\definecolor{shadecolor}{rgb}{0,0,0}%
\definecolor{greyout}{rgb}{0.65,0.7,0.7}%
\definecolor{tbhead}{rgb}{0.97,0.97,0.97}%
\definecolor{a}{rgb}{0.9,0.95,0.95}%
\definecolor{b}{rgb}{0.99,0.99,0.99}%
\definecolor{white}{rgb}{1, 1, 1}%

\newcommand{\Hair}{\ifmmode\mskip1mu\else\kern0.08em\fi}

\newcommand{\Warp}		{{Warp}\xspace} 

\begin{abstract}
We present \textit{\Warp}, a hardware platform to support research
in approximate computing, sensor energy optimization, and
energy-scavenged systems.  \Warp incorporates 11 state-of-the-art
sensor integrated circuits, computation, and an energy-scavenged
power supply, all within a miniature system that is just
3.6\,cm$\times$3.3\,cm$\times$0.5\,cm.  \Warp's sensor integrated
circuits together contain a total of 21 sensors with a range of
precisions and accuracies for measuring eight sensing modalities
of acceleration, angular rate, magnetic flux density (compass
heading), humidity, atmospheric pressure (elevation), infrared
radiation, ambient temperature, and color.  \Warp uses a combination
of analog circuits and digital control to facilitate further tradeoffs
between sensor and communication accuracy, energy efficiency, and
performance.  This article presents the design of \Warp and presents
an evaluation of our hardware implementation. The results show how
\Warp's design enables performance and energy efficiency versus
accuracy tradeoffs.
\end{abstract}

\dates{This manuscript was compiled on \today}

\begin{document}

\verticaladjustment{-5pt}

\maketitle
\thispagestyle{firststyle}
\ifthenelse{\boolean{shortarticle}}{\ifthenelse{\boolean{singlecolumn}}{\abscontentformatted}{\abscontent}}{}

\vspace{-0.1in}
\dropcap{S}ensor integrated circuits are critical components
of many hardware platforms, from augmented reality and wearable
health monitors, to drones.  Sensors convert physical signals such
as temperature, vibration, rotation, and so on, into signals which
are then digitized and used in computations.  Because sensor circuits
are often constrained by the physics of the phenomena they are
designed to measure, sensors often do not benefit from the scaling
of semiconductor technology that has enabled dramatic reduction in
power dissipation of digital logic.  As a result, sensors today
constitute an important component of the power dissipation in many
energy-constrained platforms such as those that operate of scavenged
energy.

The power dissipated by sensors depends on their electrical
configuration (e.g., supply voltage) as well as on their software
configuration (e.g., number of bits per sample for sensors with
digital interfaces).  These configuration parameters also affect
the precision and accuracy of samples produced by sensors.  System
designers can capitalize on this observation to trade energy
efficiency and performance for precision (fineness of measurement
resolution) and accuracy (difference between measured signal values
and the true signal value). These tradeoffs have been investigated
by several research efforts in the last decade~\cite{Carbin:2013,
Bornholt2014,Sampson:2011,Miguel:2014:LVA:2742155.2742169,
Darulova:2017:TCR:3062396.3014426, kim2017axserbus, lee2017high,
Hoffmann:2011:DKR:1950365.1950390, Esmaeilzadeh:2012, 189934,
EDA-049, Lingamneni:2012:AMU:2212908.2212912,
Stanley-Marbell:2016:RSI:2897937.2898079, stanley2015efficiency,
StanleyMarbell:pmup06, StanleyMarbell:itw09}.  Despite the significant
research interest in efficiency versus precision and accuracy
tradeoffs however, no common open hardware platforms for research
evaluation exist today.  This article introduces \textit{\Warp},
an open\footnote{We plan to make the hardware design files, our
basic firmware, as well as examples of measurement data, available
on GitHub.} hardware platform for evaluating hardware and software
techniques that trade precision, accuracy, and reliability for
improved efficiency in energy-constrained systems. \Warp fills an
unmet need for research evaluation hardware, and the measurements
from platforms such as \Warp could serve as valuable error models
for research on algorithms, programming languages, and software.

\begin{figure}[t!]
\centering
\includegraphics[trim=0.0cm 0.0cm 0.0cm 0.0cm, clip=true, angle=0, width=0.495\textwidth]{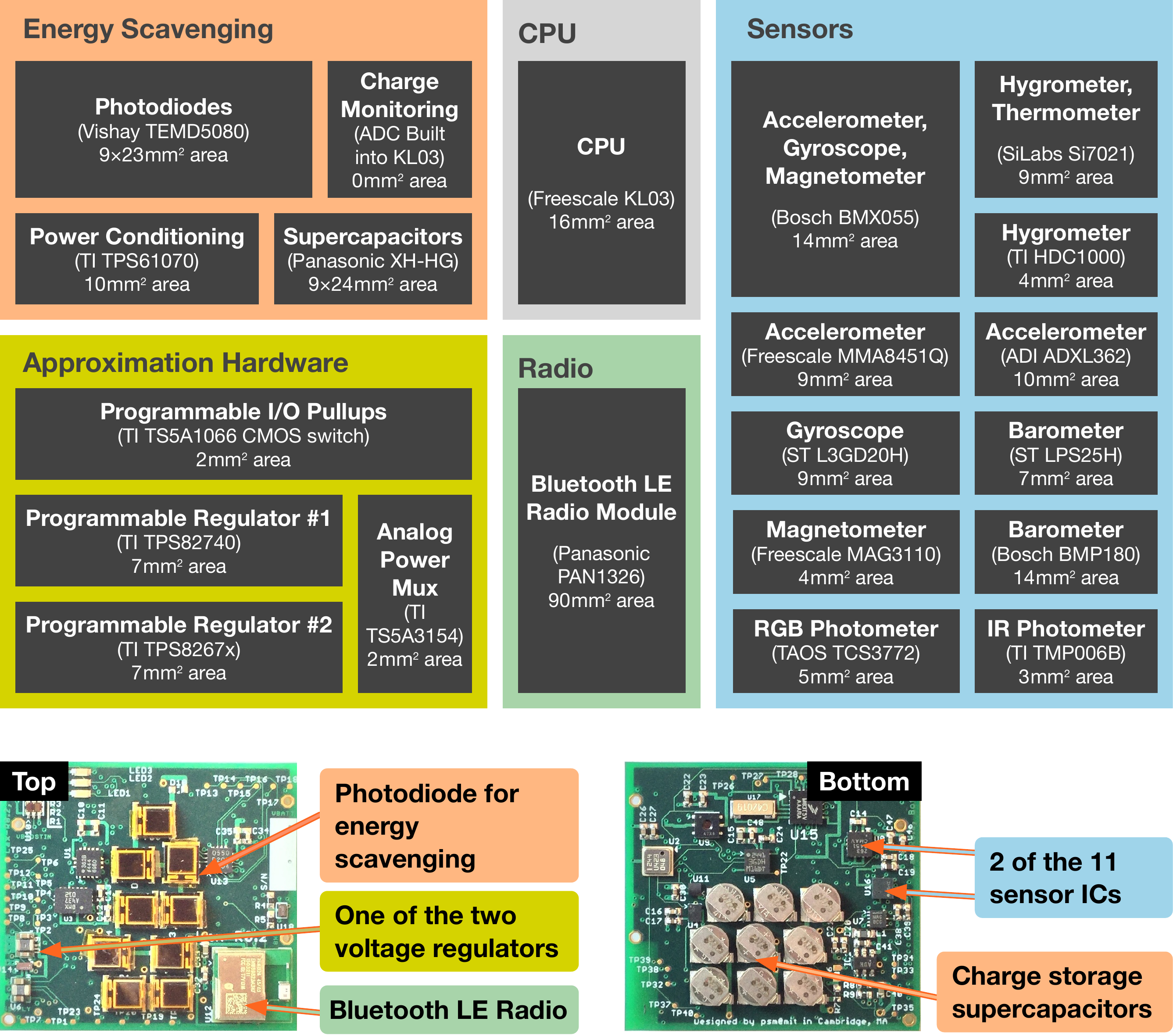}
\vspace{-0.1in}
\caption{The \Warp hardware platform contains 11 sensor integrated
circuits which together provide 21 sensors across eight sensing
modalities (see Table~\ref{table:voltages}). \Warp uses this diversity
of sensors together with circuit support for tradeoffs between precision, accuracy, performance, and
energy-efficiency.  An energy-scavenging photovoltaic subsystem
powers the platform.}
\iftoggle{warpTightFormatting}{\vspace{-0.1in}}{}
\label{fig:WarpArch}
\end{figure}

\iftoggle{warpTightFormatting}{\vspace{-0.1in}}{}
\section{The \Warp Hardware Platform}
\label{section:WarpHW}
\iftoggle{warpTightFormatting}{\vspace{-0.05in}}{}
We designed \Warp to provide a greater range of energy and correctness
tradeoffs than is available using commercially-available off-the-shelf
platforms. \Warp achieves this greater flexibility by integrating
sensors having a range of hardware-implemented precisions and
accuracies, by incorporating new hardware facilities for approximation
(Section~\ref{section:WarpHW-approximation}), and by powering the
entire platform from scavenged energy using a miniature photovoltaic
array (Section~\ref{section:WarpHW-scavenging}).  Integrating these
facilities into 3.6\,cm$\times$3.3\,cm$\times$0.5\,cm
(Section~\ref{section:WarpHW-miniaturization}) makes it feasible
to deploy \Warp in realistic use cases.  Figure~\ref{fig:WarpArch}
shows the system organization of \Warp.

\begin{table*}
\centering
\caption{Operating voltage ranges together with precision, accuracy,
and noise properties of the sensor integrated circuits in \Warp.
There are 11 sensor integrated circuits, each measuring one or more
of eight physical phenomena, for a total of 21 sensors.  Many
of \Warp's sensor integrated circuits also include software-accessible
temperature sensors (for calibration).}
\iftoggle{warpTightFormatting}{\vspace{-0.1in}}{}
\label{table:voltages}
\small
\begin{tabular}{lllll}
\toprule
\textbf{Sensor}						& \textbf{Supply}		& \textbf{Accuracy Range}						& \textbf{Interface Precision}	& \textbf{Significant Digits and Coefficient}\\
									& \textbf{Range}		& 												& 								& \textbf{of Variation for 100 z-axis }\\
									& \textbf{(V)}			& \textbf{(Noise Measure)}						& \textbf{(bits/sample)}		& \textbf{Measurements Performed at 2.5V.}\\
\hline
\rowcolor{a}MMA8451Q accelerometer	& 1.95 -- 3.6			& 99 -- 126\,$\mu$g/$\sqrt{\mathrm{Hz}}$		&	8 or 14						& 0 digits$_{10}$, 1 digits$_{2}$, COV:  94.1\%\\
\rowcolor{b}BMX055 accelerometer	& 2.4 -- 3.6			& 150\,$\mu$g/$\sqrt{\mathrm{Hz}}$				&	8 or 12						& 1 digits$_{10}$, 4 digits$_{2}$, COV:  0.9\%\\
\rowcolor{a}ADXL362 accelerometer	& 1.6 -- 3.5			& 175 -- 550\,$\mu$g/$\sqrt{\mathrm{Hz}}$		&	4, 8, or 12					& 0 digits$_{10}$, 1 digits$_{2}$, COV:  5.6\%\\
\hline
\rowcolor{b}L3GD20H gyroscope		& 2.2 -- 3.6			& 0.011\,\textdegree/s/$\sqrt{\mathrm{Hz}}$		&	8 or 16						& 0 digits$_{10}$, 1 digits$_{2}$, COV:  115.8\%\\
\rowcolor{a}BMX055 gyroscope		& 2.4 -- 3.6			& 0.014\,\textdegree/s/$\sqrt{\mathrm{Hz}}$		&	8 or 16						& 0 digits$_{10}$, 1 digits$_{2}$, COV:  118.1\%\\
\hline
\rowcolor{b}MAG3110 magnetometer	& 1.95 -- 3.6			& 0.25 -- 0.4\,$\mu$T							&	8 or 16						& 1 digits$_{10}$, 3 digits$_{2}$, COV:  1.3\%\\
\rowcolor{a}BMX055 magnetometer		& 2.4 -- 3.6			& 0.3 -- 1.4\,$\mu$T							&	8 or 13 ($x$-, $y$-), 15 ($z$-)& 1 digits$_{10}$, 1 digits$_{2}$, COV:  2.2\%\\
\hline
\rowcolor{b}SI7021 hygrometer		& 1.9 -- 3.6			& $\pm$2\% accuracy								&	8, 10, 11, or 12			& ---\\
\rowcolor{b}						& 						& $\pm$0.025--0.2\% precision					&								& \\
\rowcolor{a}HDC1000 hygrometer		& 3.0 -- 5.0			& $\pm$4\% accuracy								&	14							& ---\\
\rowcolor{a}						& 						& $\pm$0.1\% precision							&								& \\
\hline
\rowcolor{b}LPS25H barometer		& 1.7 -- 3.6			& 0.01 -- 0.03\,hPa								&	8, 16, or 24				& ---\\
\rowcolor{a}BMP180 barometer		& 1.6 -- 3.6			& 0.03 -- 0.06\,hPa								&	8, 16, or 19				& ---\\
\hline
\rowcolor{b}HDC1000 thermometer		& 3.0 -- 5.0			& $\pm$0.2\textdegree C							&	14							& ---\\
\rowcolor{a}SI7021 thermometer		& 1.9 -- 3.6			& $\pm$0.3\textdegree C							&	11, 12, 13, or 14			& ---\\
\rowcolor{b}ADXL362 thermometer		& 1.6 -- 3.5			& $\pm$0.5\textdegree C							&	4 or 12						& ---\\
\rowcolor{a}TMP006B thermometer		& 2.2					& $\pm$1\textdegree C							&	8 or 14						& ---\\
\rowcolor{b}BMP180 thermometer		& 1.6 -- 3.6			& $\pm$1\textdegree C							&	8 or 16						& ---\\
\rowcolor{a}MAG3110 thermometer		& 1.95 -- 3.6			& $\ge \pm$1\textdegree							&	8							& ---\\
\rowcolor{b}L3GD20H thermometer		& 2.2 -- 3.6			& $\ge \pm$1\textdegree							&	8							& ---\\
\rowcolor{a}LPS25H thermometer		& 1.7 -- 3.6			& $\pm$2\textdegree C							&	8 or 16						& ---\\
\rowcolor{b}BMX055 thermometer		& 2.4 -- 3.6			& $\pm$2\textdegree C							&	8							& ---\\
\hline
\rowcolor{a}TCS3772	photometer		& 2.7 -- 3.3			& 14\%--35\% Irradiance Resp.					&	8 or 16 per R/G/B/clear		& ---\\
\bottomrule
\end{tabular}
\normalsize
\iftoggle{warpTightFormatting}{\vspace{-0.2in}}{}
\end{table*}

The sensors in \Warp cover eight sensing modalities: \ding{202}
temperature, \ding{203} acceleration in three axes, \ding{204}
angular rate in three axes, \ding{205} magnetic flux density in
three axes (often used as a digital compass), \ding{206} humidity,
\ding{207} pressure (for measuring, e.g., atmospheric pressure or
elevation), \ding{208} infrared (IR) radiation, and \ding{209} color
(a red-green-blue-clear sensor with filters for 615\,nm, 525\,nm,
and 465\,nm light). For each of the first six modalities, \Warp
contains at least two different state-of-the-art sensor hardware
implementations, each of which represents a different point in the
tradeoff space between precision, accuracy, power dissipation, and
performance. For example, for atmospheric pressure, \Warp contains
both an LPS25H integrated circuit (IC) as well as a BMP180 IC.
For each sensing modality, the sensors provide
a range of accuracies and precision as specified by their datasheets
(i.e., Type B uncertainty~\cite{kirkup2006introduction}).
Table~\ref{table:voltages} lists the sensors, their operating voltage
ranges, and appropriate metrics for characterizing their output precision,
accuracy, and noise. 

The sensors in Table~\ref{table:voltages} have a range of accuracies
and output noise. By including multiple hardware implementations
of sensors for the same sensing modality. \Warp allows its users
to evaluate techniques that tradeoff accuracy for power.  Having
multiple sensors for the same modality also makes it possible to
implement techniques for improving accuracy, such as using simultaneous
sampling together with signal correlation to improve signal-to-noise
ratios in sensing applications.

For the accelerometers, gyroscopes, and magnetometers in \Warp, the
last column in Table~\ref{table:voltages} quantifies the number of
measurement value digits that remain stable in a set of 100
measurements conducted with the measurand nominally unchanging.
These results show that the noise inherent in measurements using
the sensors in Table~\ref{table:voltages} as represented by the
number of significant digits or the coefficient of variation (COV)
is consistently large: In all the measurements, the ratio of the
standard deviation to the mean (i.e., the coefficient of variation
(COV)) is at least 0.9\% and in some cases over 110\%. Because of
this variation, the impact of sensor data approximation techniques
which introduce random errors in sensor values~\cite{189934} or
which introduce quantization noise without decreasing a signal's
dynamic range~\cite{Stanley-Marbell:2016:RSI:2897937.2898079} may
be naturally masked by inherent measurement uncertainty which the
applications which consume sensor data must already contend with.
Because its sensors allow sampling at a range of bits per sample
(column 4 of Table~\ref{table:voltages}), \Warp also supports
applications which may be intolerant of noise, but which can tolerate
reduced precision or reduced dynamic range.

\iftoggle{warpTightFormatting}{\vspace{-0.1in}}{}
\subsection{Hardware support for approximate sensing}
\label{section:WarpHW-approximation}
\iftoggle{warpTightFormatting}{\vspace{-0.05in}}{}
\Warp implements two complementary hardware facilities for trading
improved sensor energy efficiency for sensor data precision, sensor
data accuracy, sensor access reliability, and sensor measurement
latency.

{\bf Sensor accuracy and reliability tradeoffs:} \Warp implements
the Lax~\cite{189934} sensor hardware approximation technique using
two software-controlled voltage regulators. \Warp uses two miniature
voltage regulators, both occupying less than 7\,mm$^2$ in circuit
board area. The \Warp printed circuit board design can be populated
with one regulator with a programmable output voltage range of
either 1.8\,V to 2.5\,V in steps of 0.1\,V or 2.6\,V to 3.3\,V in
steps of 0.1\,V, and a second regulator with a fixed output voltage
of either 1.05\,V, 1.1\,V, 1.2\,V, 1.225\,V, 1.26\,V, 1.5\,V, 1.6\,V,
1.8\,V, 1.86\,V, 1.95\,V, or 2.1\,V. The outputs of these two
regulators are fed into a software-controlled analog switch. Using
this configuration, \Warp can control sensor supply voltages to be
any value in the range of 1.8\,V to 2.5\,V or 2.6\,V to 3.3\,V in
steps of 0.1\,V, together with the additional voltages listed above
which have to be fixed when the components are mounted on the circuit
board. \Warp's sensor supply voltage changes have a typical hardware
latency of 315\,$\mu$s due to the output voltage switching latency
of the voltage regulators and the switching time of the analog
switch.

{\bf Sensor communication reliability tradeoffs:} \Warp implements
a hardware facility to allow software control of the pull-up resistors
which are mandatory for the I2C serial communication standard used
by most sensor integrated circuits.  Disabling the I/O pull-up
reduces the reliability of communication, but removes the main
source of power dissipation for open-drain interfaces such as I2C.
For an I2C interface operating at an I/O supply voltage of 2.5\,V,
the average power dissipated in a 4.7\,k$\Omega$ pull-up resistor is 1.3\,mW;	%
this is more than the power dissipation of most sensors in \Warp.

\iftoggle{warpTightFormatting}{\vspace{-0.1in}}{}
\subsection{Energy scavenging}
\label{section:WarpHW-scavenging}
\iftoggle{warpTightFormatting}{\vspace{-0.05in}}{}
The \Warp platform contains a photodiode array for energy scavenging.
A boost regulator (TI TPS61070) serves as a charge pump between the
series connected photodiodes and a supercapacitor array having a total
capacitance of 0.72\,F.

\iftoggle{warpTightFormatting}{\vspace{-0.1in}}{}
\subsection{Implementation miniaturization}
\label{section:WarpHW-miniaturization}
\iftoggle{warpTightFormatting}{\vspace{-0.05in}}{}
We optimized the implementation of \Warp for size to achieve a form
factor that is small enough for use in user studies (e.g., as a
wearable platform). To achieve this level of integration, we
implemented \Warp using a 10-layer printed circuit board process
with a board thickness of 62\,mils (1.6\,mm). Fully populated with
components, the \Warp prototype is only \textasciitilde5\,mm thin.
The platform is programmed via an auxiliary extension of the main
implementation board which contains a second processor (identical to
the main processor).

\iftoggle{warpTightFormatting}{\vspace{-0.1in}}{}
\section{Evaluation}
\label{section:evaluation}
\iftoggle{warpTightFormatting}{\vspace{-0.05in}}{}
We use a Keysight B2962A source-measure unit (SMU) for power
measurements. The B2962A is a laboratory-grade 6.5-digit 4-quadrant
SMU intended for low-power circuit characterization. It provides
current sourcing precision of 10\,fA, voltage sourcing precision
of 100\,nV, current measurement precision of 10\,nA, and voltage
measurement precision of 200\,mV. These current and voltage measurement
specifications enable us to measure power dissipation to a resolution
of better than 1\,$\mu$W.

\begin{figure}[t]
\centering
\subfloat[]{\includegraphics[trim=0.0cm 0.0cm 0.0cm 0.0cm, clip=true, angle=0, width=0.24\textwidth]{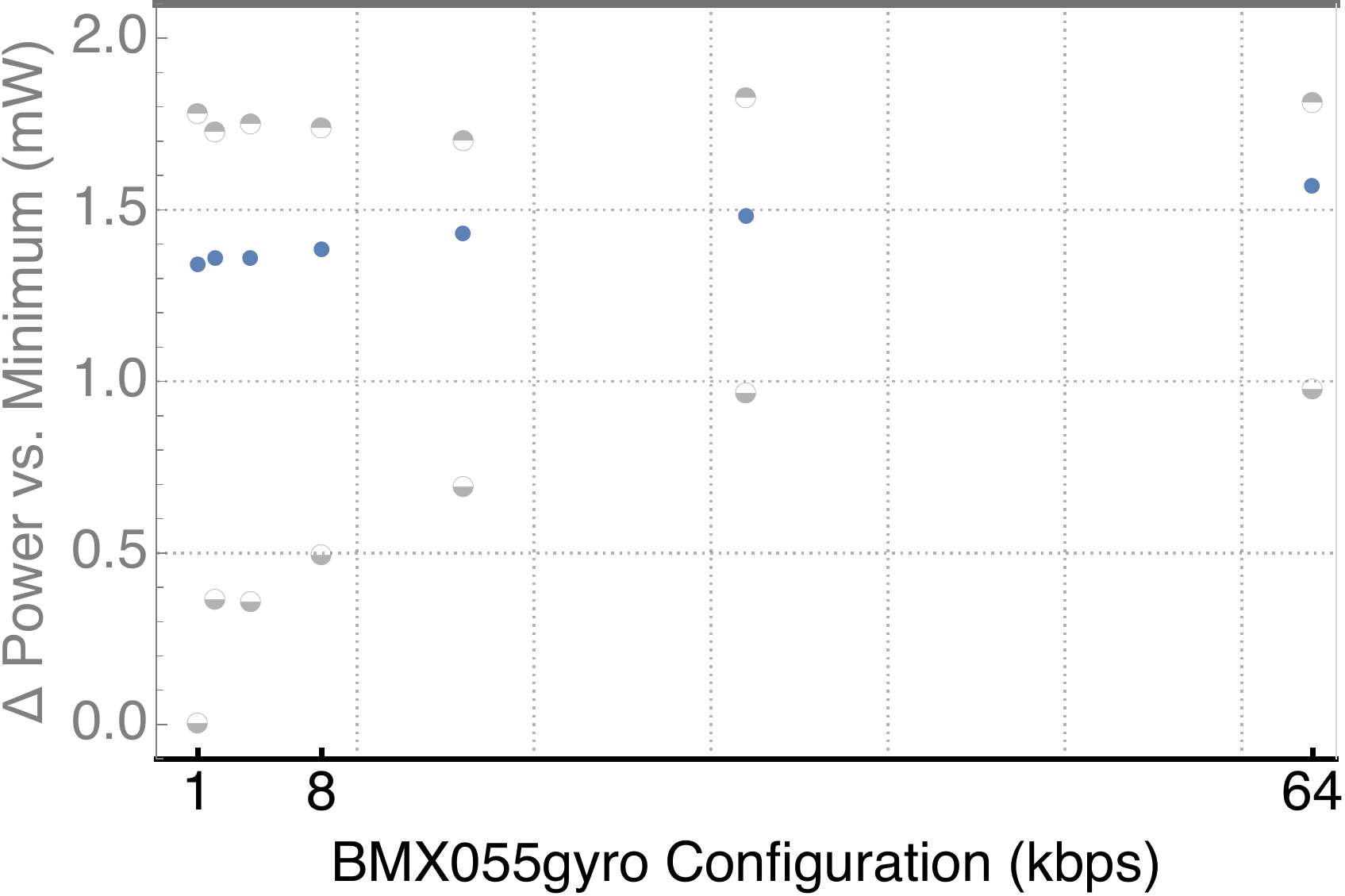}}~
\subfloat[]{\includegraphics[trim=0.0cm 0.0cm 0.0cm 0.0cm, clip=true, angle=0, width=0.24\textwidth]{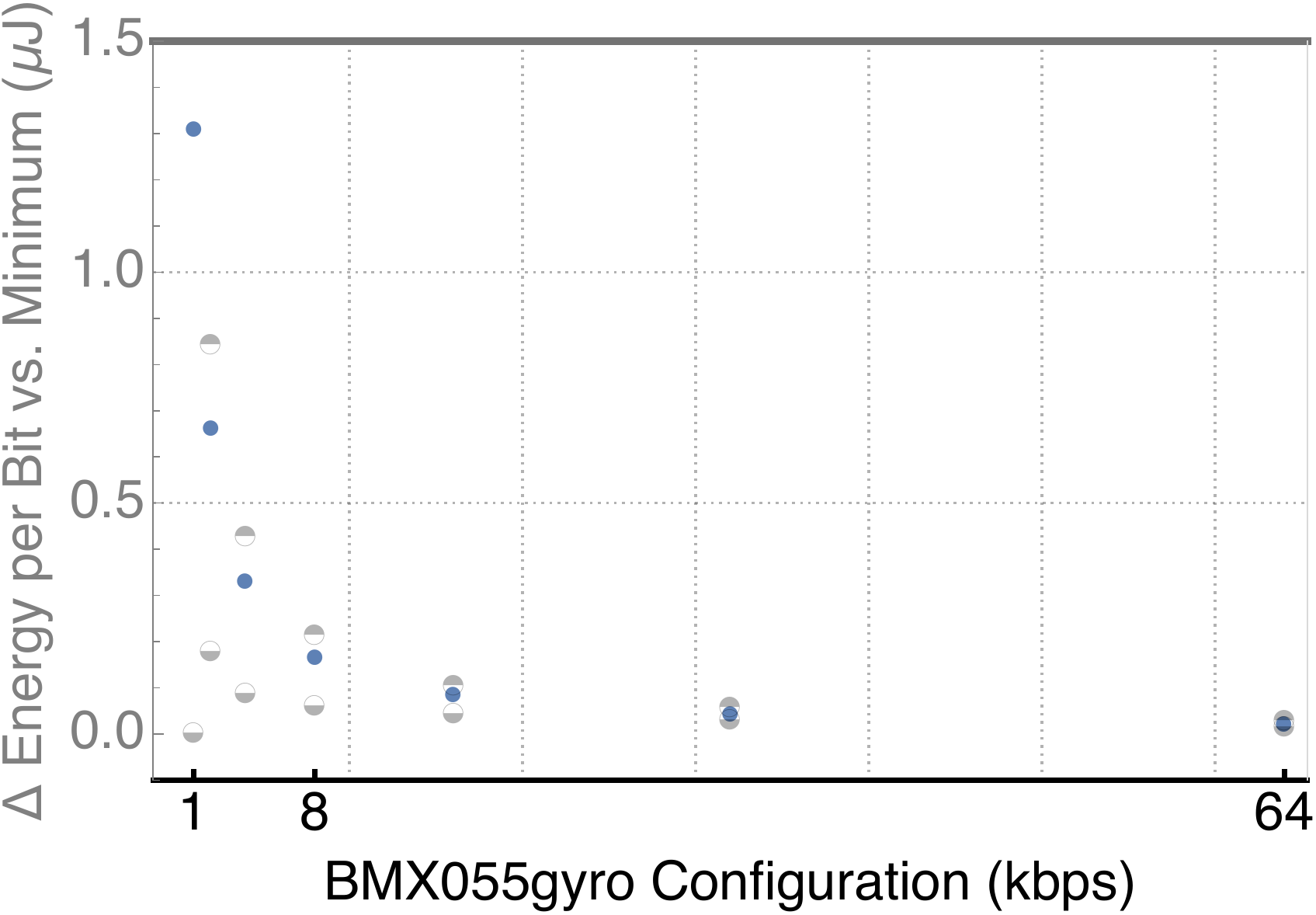}}\\
\subfloat[]{\includegraphics[trim=0.0cm 0.0cm 3.5cm 0.0cm, clip=true, angle=0, width=0.21\textwidth]{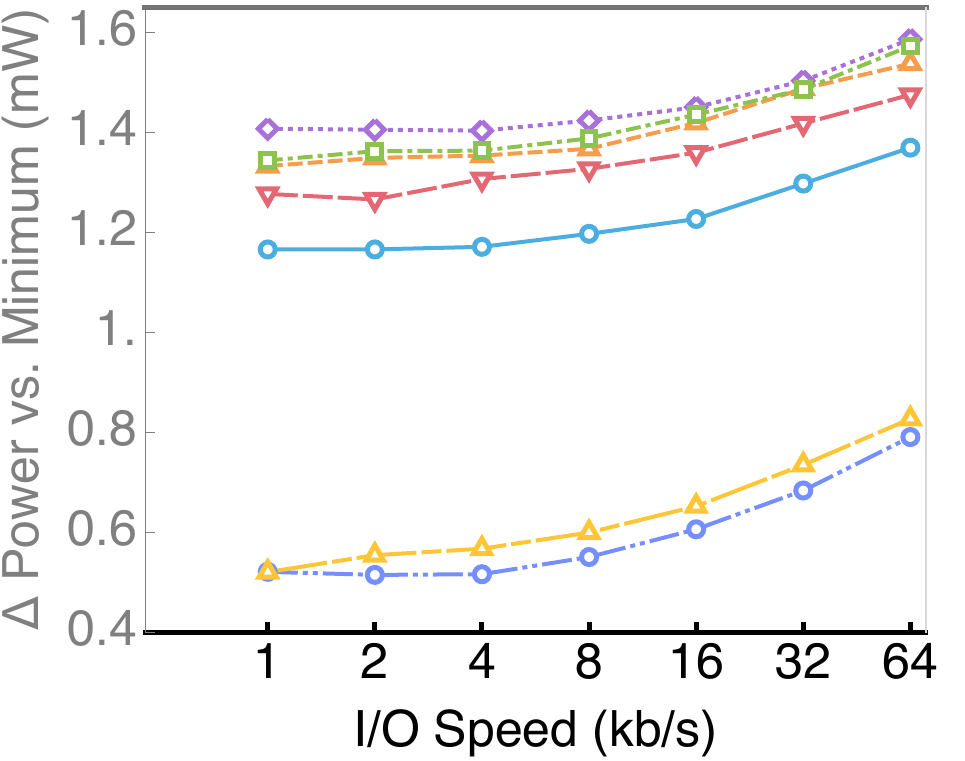}}
\subfloat[]{\includegraphics[trim=0.0cm 0.0cm 0.0cm 0.0cm, clip=true, angle=0, width=0.285\textwidth]{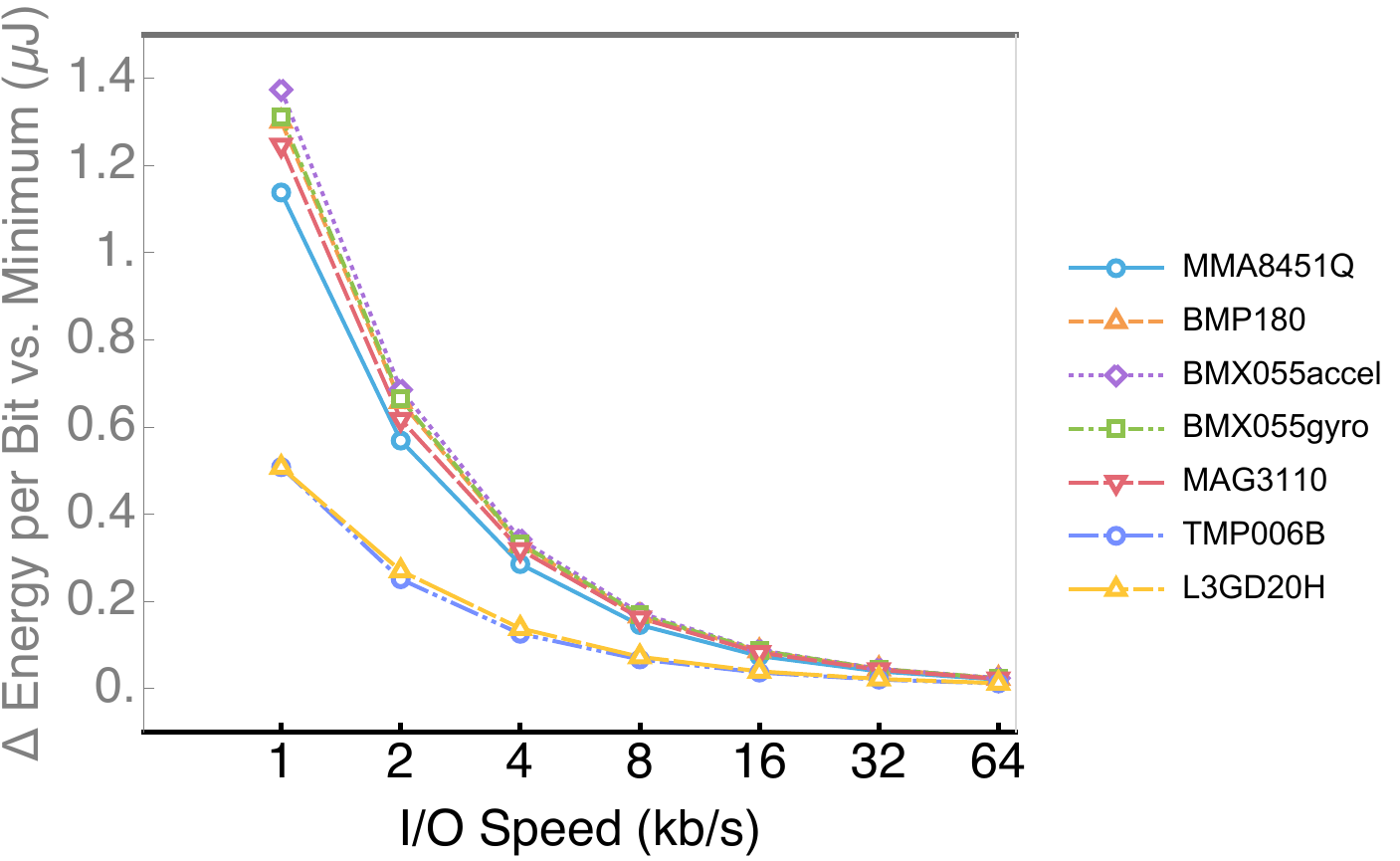}}
\iftoggle{warpTightFormatting}{\vspace{-0.1in}}{}
\caption{\Warp enables tradeoffs between I/O power dissipation, energy per bit, and I/O data transfer speeds.}
\iftoggle{warpTightFormatting}{\vspace{-0.1in}}{}
\label{fig:results-issue-33-power-versus-all-bitrate-deltaMeansPlot}
\end{figure}

\iftoggle{warpTightFormatting}{\vspace{-0.1in}}{}
\subsection{Performance versus power tradeoff results}
\label{section:power-results}
\iftoggle{warpTightFormatting}{\vspace{-0.05in}}{}
Figure~\ref{fig:results-issue-33-power-versus-all-bitrate-deltaMeansPlot}(a)
shows a representative example of how the power dissipation for
accessing a sensor (the BMX055 gyroscope) varies with I/O speed.
For the BMX055 gyroscope, power dissipation increases by over
0.2\,mW as the speed at which the sensor is accessed is increased
from 1\,kb/s to 64\,kb/s. Even though power dissipation increases
with I/O speed, Figure~\ref{fig:results-issue-33-power-versus-all-bitrate-deltaMeansPlot}(b)
shows that the energy per bit for I/O decreases exponentially with
I/O speed.

Figure~\ref{fig:results-issue-33-power-versus-all-bitrate-deltaMeansPlot}(c)
and Figure~\ref{fig:results-issue-33-power-versus-all-bitrate-deltaMeansPlot}(d)
show similar trends in I/O power and energy per bit for seven of \Warp's
sensors and shows how power dissipation varies by 0.2\,mW\,--\,0.3\,mW
as a function of I/O speed. The magnitude of this change in I/O
power dissipation is greater than the power dissipation of many of
the sensors in the platform, motivating the need for precise and
approximate techniques for improving I/O power efficiency.

\begin{figure}
\centering
\subfloat[Distributions of $z$-axis magnetic flux for BMX055 operating at supply voltages from 1.8\,V to 2.5\,V.]{\includegraphics[trim=0.0cm 0.0cm 0.0cm 0.0cm, clip=true, angle=0, width=0.235\textwidth]{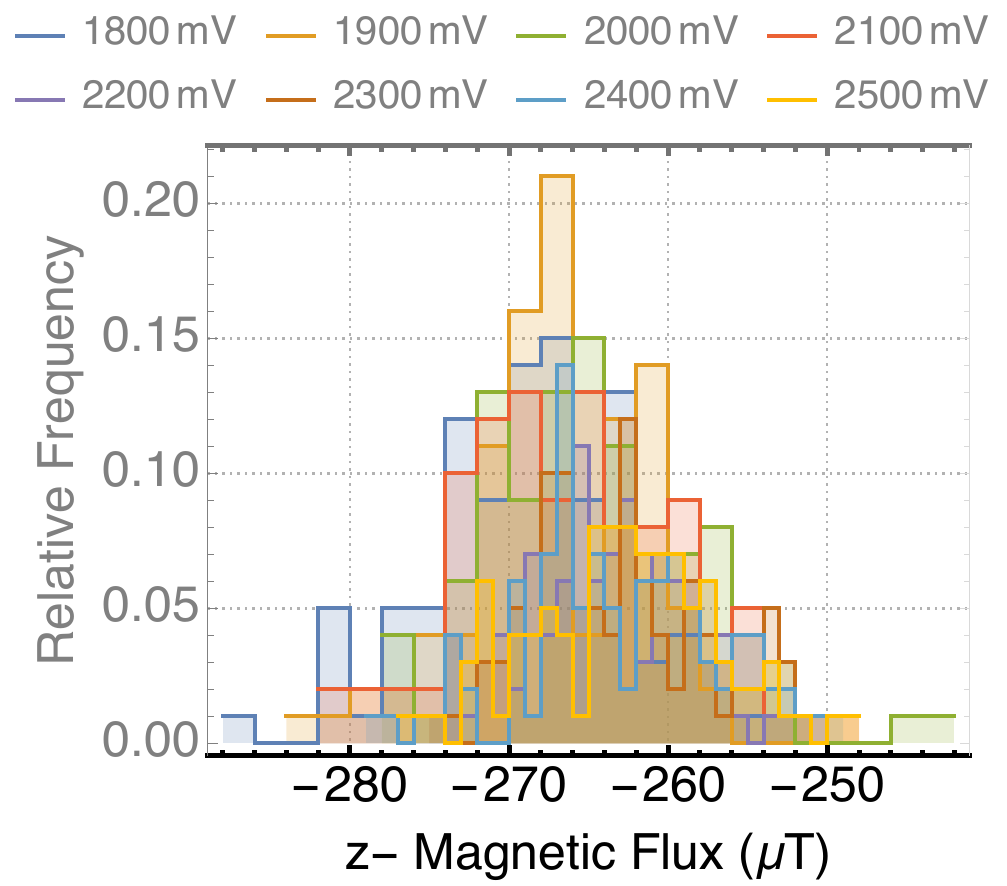}}\hspace{0.1in}
\subfloat[100 measurements of $z$-axis magnetic flux for BMX055 at 2.2\,V. Passes normality test (Gaussian overlaid).]{\includegraphics[trim=0.0cm 0.0cm 0.0cm 4.80cm, clip=true, angle=0, width=0.23\textwidth]{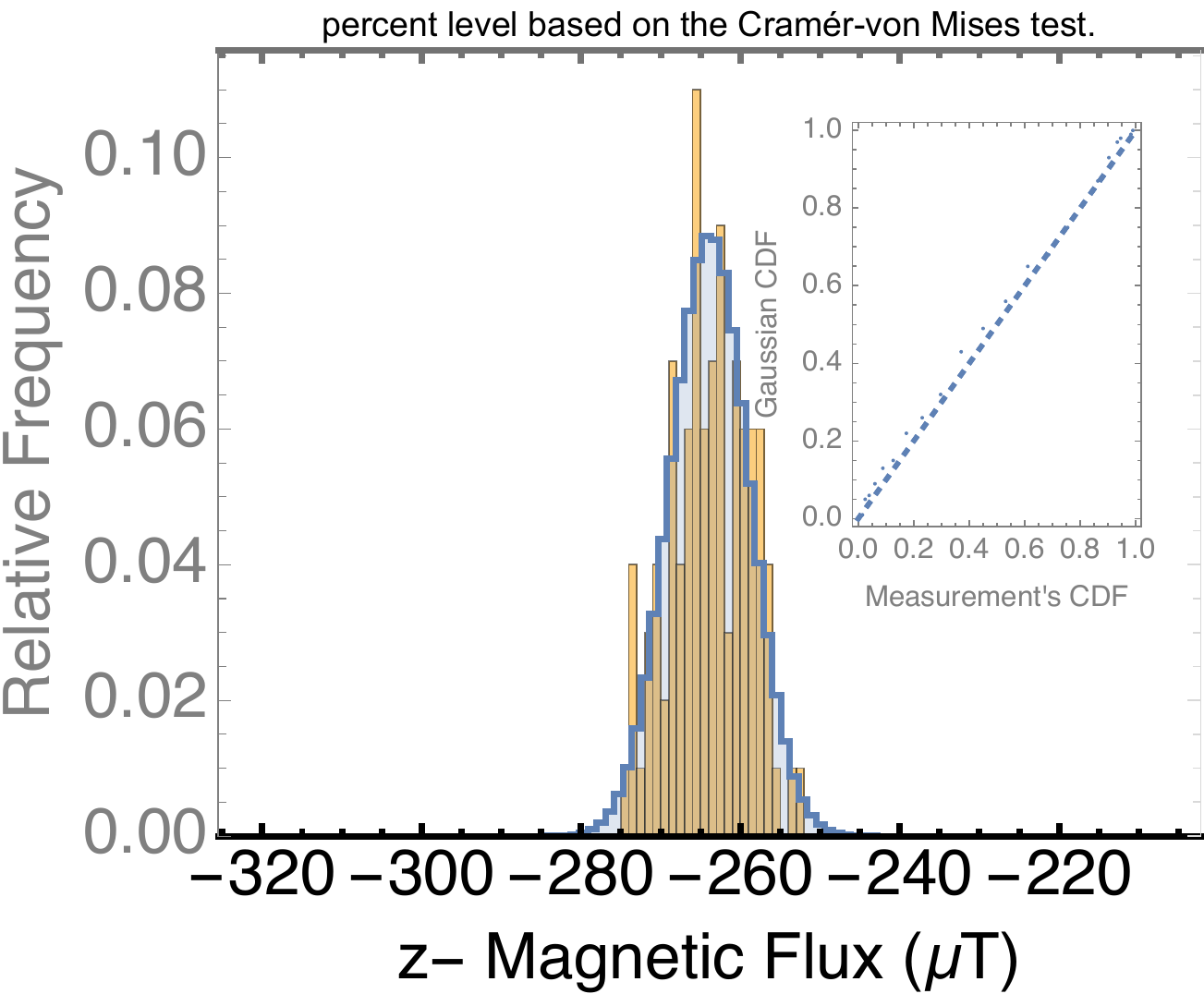}}\\
\subfloat[Distribution of $y$-axis acceleration for ADXL362 operating at supply voltages from 1.8\,V to 2.5\,V.]{\includegraphics[trim=0.0cm 0.0cm 0.0cm 0.0cm, clip=true, angle=0, width=0.235\textwidth]{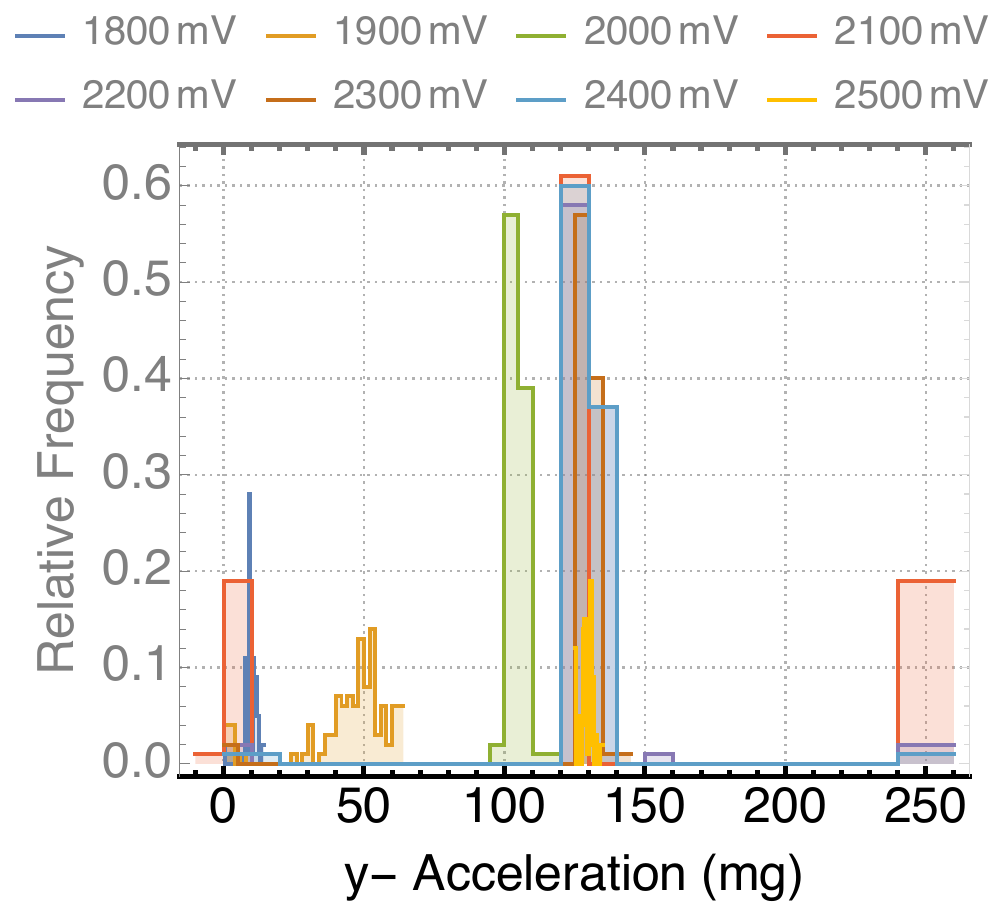}}\hspace{0.1in}
\subfloat[100 measurements of $y$-axis acceleration for ADXL362 at 2.2\,V. Fails normality test (Gaussian overlaid).]{\includegraphics[trim=0.0cm 0.0cm 1.85cm 5.35cm, clip=true, angle=0, width=0.23\textwidth]{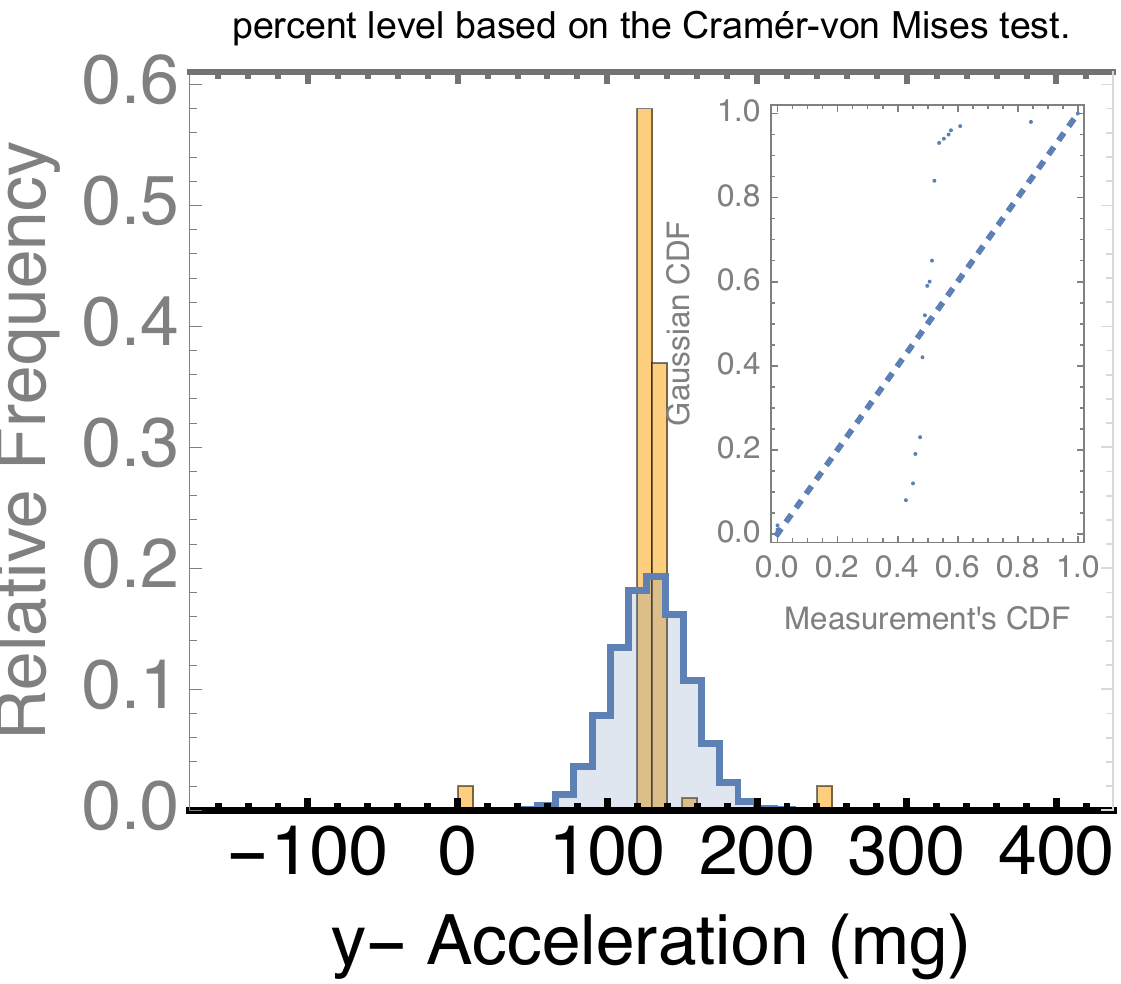}}
\iftoggle{warpTightFormatting}{\vspace{-0.05in}}{}
\caption{Distributions of sensor readings differ across sensor modalities and across integrated circuit implementations and vary with supply voltage.}
\iftoggle{warpTightFormatting}{\vspace{-0.2in}}{}
\label{fig:results-outlineHistogram-for-all-voltages}
\end{figure}

\iftoggle{warpTightFormatting}{\vspace{-0.1in}}{}
\subsection{Sensor accuracy versus voltage tradeoff results}
\label{section:approximation-results}
\iftoggle{warpTightFormatting}{\vspace{-0.05in}}{}
We evaluate the tradeoff between accuracy of sensor data, power,
and performance, by operating all three of the accelerometers and
two of the gyros in \Warp over a range of supply voltages. For each
of the three axes of these five sensors (15 signal dimensions
in total), we operate the sensors at one of eight supply voltages
uniformly spaced between 1.8\,V and 2.5\,V, a total of 120 measurement
configurations. In each of these 120 measurement configurations,
we compare the average of 100 sensor signal measurements at each
of the eight supply voltage settings to an average of 100 sensor
measurements when the sensor is operating under identical conditions
but at a supply voltage of 2.5\,V.
Figure~\ref{fig:results-outlineHistogram-for-all-voltages} shows
examples of the distributions of values from two of the 15 signal
types. Figure~\ref{fig:results-outlineHistogram-for-all-voltages}(a)
shows the distributions of z-axis magnetic flux values returned by
the BMX055 magnetometer (typically used in consumer applications
as a digital compass), in a fixed orientation, as we change the
supply voltage of the sensor from 1.8\,V to 2.5\,V.
Figure~\ref{fig:results-outlineHistogram-for-all-voltages}(b) shows
the distribution of sensor values measured at 2.5\,V, with a histogram
of random variates drawn from a Gaussian distribution with the same
mean and variance overlaid.  The null hypothesis that the data is
distributed according to the Gaussian with the same mean and variance
as the sample is not rejected at the 5\% level based on the
Cram\'{e}r-von Mises test.

Figure~\ref{fig:results-outlineHistogram-for-all-voltages}(c) shows
the distributions of y-axis acceleration sensor values obtained
from the ADXL362 accelerometer in a fixed orientation, as a function
of sensor supply voltage. The distributions in
Figure~\ref{fig:results-outlineHistogram-for-all-voltages}(c) show
significantly greater separation than those in
Figure~\ref{fig:results-outlineHistogram-for-all-voltages}(a) and
are distinctly non-Gaussian, as the overlay of the Gaussian with
the same mean and variance in
Figure~\ref{fig:results-outlineHistogram-for-all-voltages}(d) shows.
The null hypothesis that the data is distributed according
to the Gaussian with the same mean and variance as the sample is rejected at the 5\%
level based on the Cram\'{e}r-von Mises test.

\begin{figure}
\centering
\subfloat{\includegraphics[trim=0.0cm 0.0cm 0.0cm 0.0cm, clip=true, angle=0, width=0.475\textwidth]{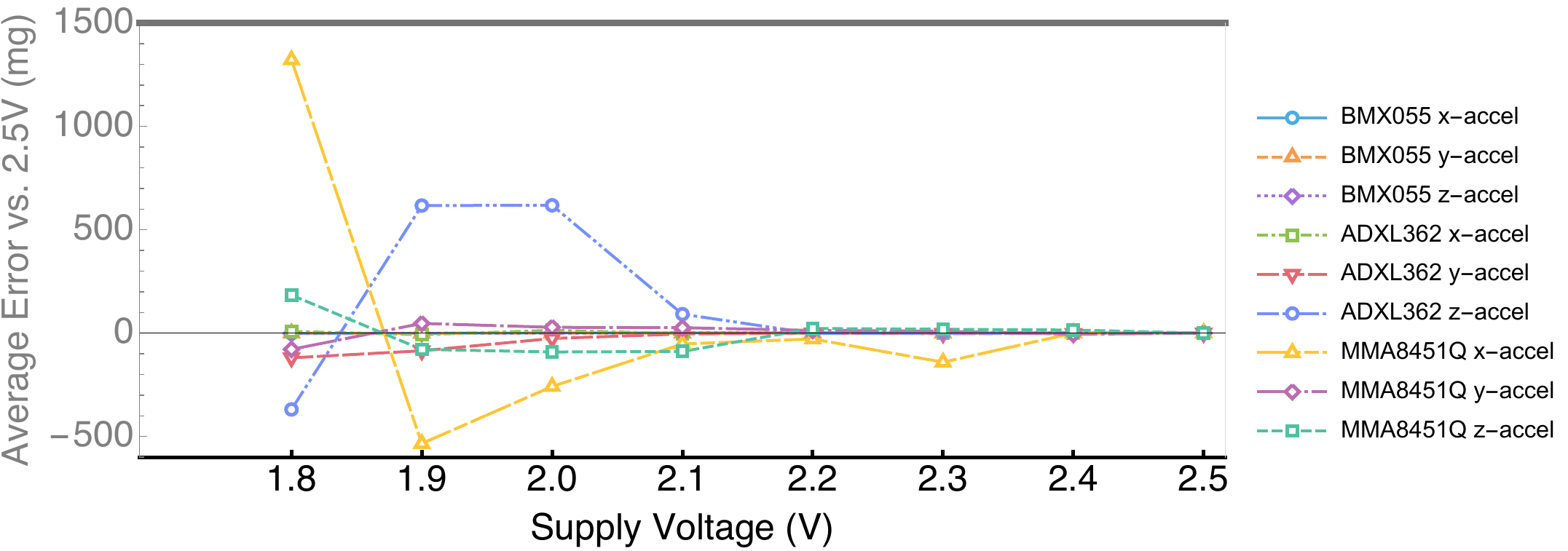}}\\
\iftoggle{warpTightFormatting}{\vspace{-0.05in}}{}
\caption{Acceleration inaccuracy (difference in value versus value
when supply voltage is at the nominal 2.5\,V). The nine data series
in the plots are acceleration readings across three axes ($x$, $y$,
and $z$) of the three accelerometers in \Warp.}
\iftoggle{warpTightFormatting}{\vspace{-0.2in}}{}
\label{fig:results-issue-39-power-versus-all-bitrate-deltaMeansPlot}
\end{figure}

\begin{figure}
\centering
\subfloat{\includegraphics[trim=0.0cm 0.0cm 0.0cm 0.0cm, clip=true, angle=0, width=0.475\textwidth]{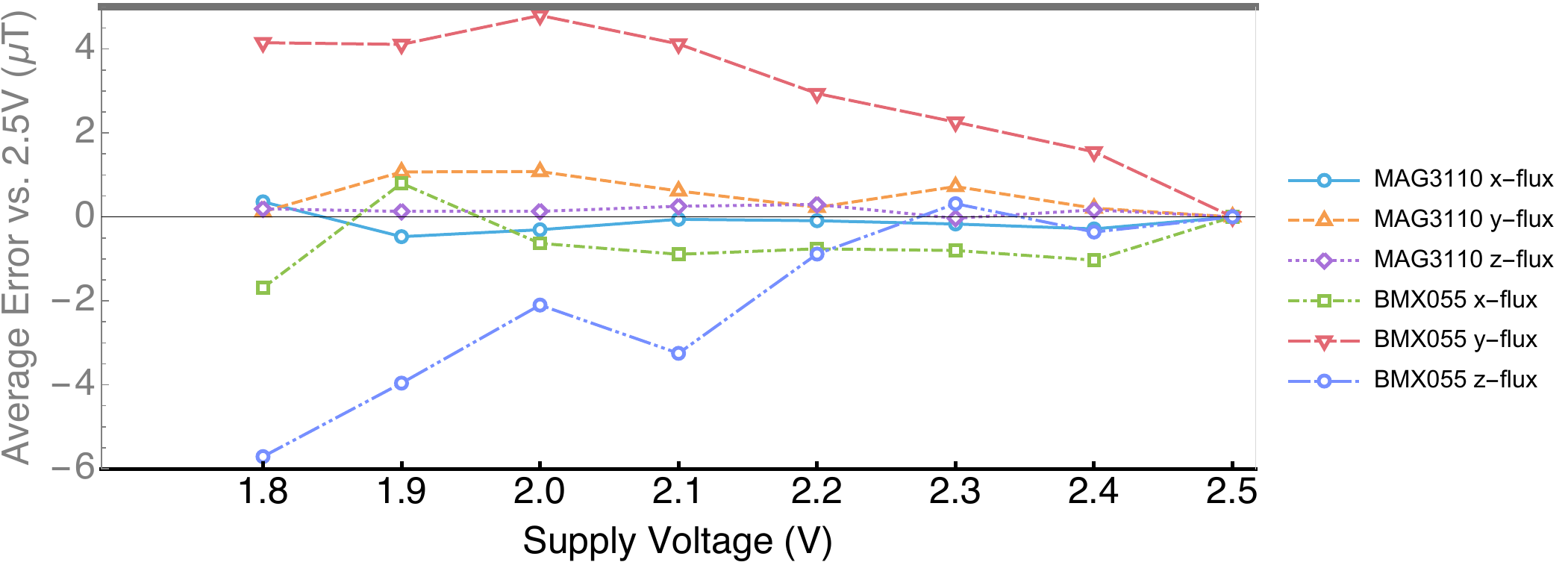}}\\
\iftoggle{warpTightFormatting}{\vspace{-0.05in}}{}
\caption{Magnetic flux inaccuracy (difference in value versus value
when supply voltage is at the nominal 2.5\,V). The six data series
in the plots are angular rate readings across three axes ($x$, $y$,
and $z$) of the two magnetometers in \Warp.}
\iftoggle{warpTightFormatting}{\vspace{-0.2in}}{}
\label{fig:results-issue-41-power-versus-all-bitrate-deltaMeansPlot}
\end{figure}

\begin{figure}
\centering
\subfloat{\includegraphics[trim=0.0cm 0.0cm 0.0cm 0.0cm, clip=true, angle=0, width=0.475\textwidth]{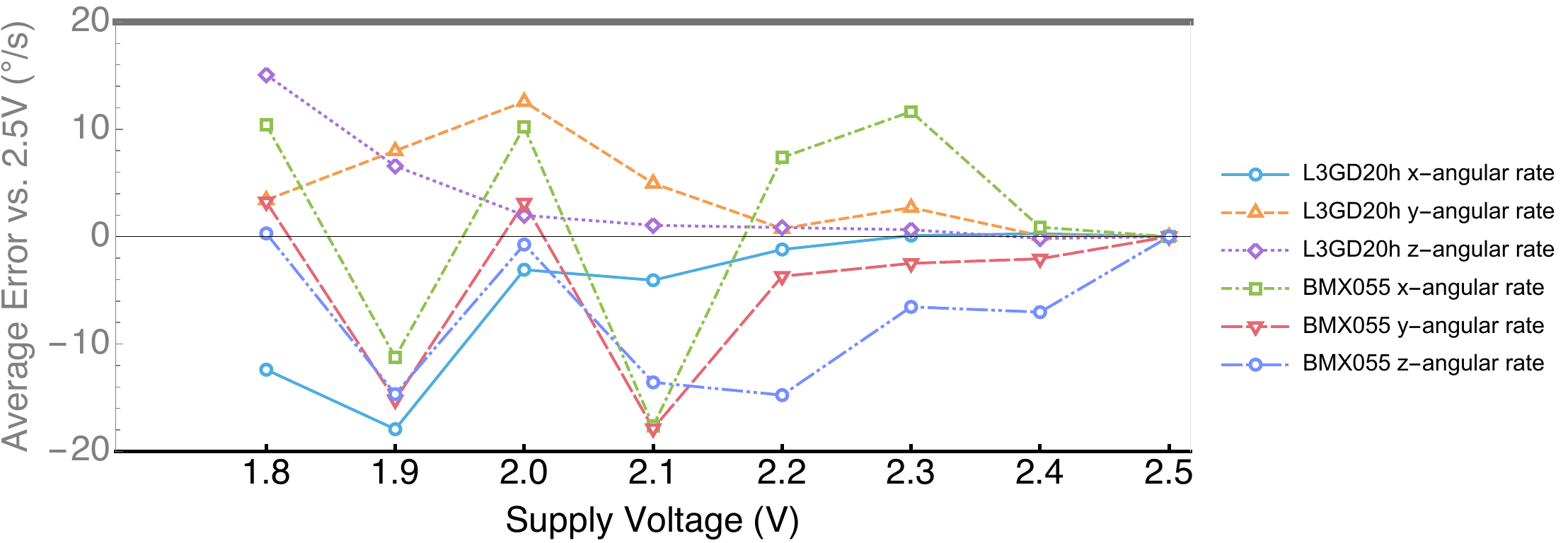}}\\
\iftoggle{warpTightFormatting}{\vspace{-0.05in}}{}
\caption{Angular rotation rate inaccuracy (difference in value versus value
when supply voltage is at the nominal 2.5\,V). The six data series
in the plots are angular rate readings across three axes ($x$, $y$,
and $z$) of the two gyroscopes in \Warp.}
\iftoggle{warpTightFormatting}{\vspace{-0.2in}}{}
\label{fig:results-issue-40-power-versus-all-bitrate-deltaMeansPlot}
\end{figure}

Figure~\ref{fig:results-issue-39-power-versus-all-bitrate-deltaMeansPlot},
Figure~\ref{fig:results-issue-41-power-versus-all-bitrate-deltaMeansPlot},
and
Figure~\ref{fig:results-issue-40-power-versus-all-bitrate-deltaMeansPlot}
show the convergence of the arithmetic mean of 100 samples taken
at each sensor operating voltage, as a function of voltage. The
results show that the accelerometers and magnetometers in \Warp
provide a useable tradeoff between supply voltage (and hence power
dissipation) and accuracy with respect to the output at a reference
operating voltage (2.5\,V in our measurements). The gyroscopes
provide less distinct trend in improving accuracy from higher supply
voltage operation. We attribute this observation to the higher
variance in the output of the gyros.  As the last column of
Table~\ref{table:voltages} shows, both the BMX055 and the L3GD20H
gyroscopes have high coefficients of variation of over 115\%,
indicating that the value of the standard deviation across the 100
samples in each measurement set was even larger than the value of
the mean.

\iftoggle{warpTightFormatting}{\vspace{-0.1in}}{}
\section{Conclusions}
\iftoggle{warpTightFormatting}{\vspace{-0.05in}}{}
Data from embedded sensing systems form the foundation for applications
ranging from wearable health monitors to infrastructure monitoring
and augmented reality.  In many of these sensor-driven systems,
energy is severely constrained and techniques to improve energy
efficiency or to trade energy efficiency for some other system
metric are valuable. This article introduces \Warp, an open hardware
platform for research in energy efficiency, performance, and
approximation tradeoffs for energy-scavenged systems. \Warp contains
a total of 21 sensors covering eight sensing modalities, a processor,
a Bluetooth Low Energy (Bluetooth LE) radio for communication, and is powered by a
photovoltaic energy scavenging array, all within a miniature system
of just 3.6\,cm$\times$3.3\,cm$\times$0.5\,cm and complements existing
research platforms targeted at precise execution on RF-scavenged energy~\cite{buettner2008rfid}.

\Warp integrates custom hardware in the form of programmable I/O
pullups and dynamically reconfigurable sensor supply voltages to
enable performance and energy efficiency versus accuracy tradeoffs.
This article presents an overview of the design of \Warp and presents
measurement results demonstrating \Warp's performance and
energy-efficiency versus accuracy tradeoffs.

\acknow{This research is supported by an Alan Turing Institute award
TU/B/000096 under EPSRC grant EP/N510129/1, and by Royal Society grant RG170136.}

\showacknow %

\pnasbreak

\bibliography{warp}

\end{document}